\documentclass[a4paper,12pt]{article}
\usepackage{jcappub}

\begin{document}
\title{Simulating neutrino echoes induced by secret neutrino interactions}

\author[1,2,3]{Jose Alonso Carpio}
\author[1,2,3,4,5]{Kohta Murase}
\affiliation[1]{Department of Physics, The Pennsylvania State University, University Park, Pennsylvania 16802, USA}
\affiliation[2]{Department of Astronomy and Astrophysics, The Pennsylvania State University, University Park, Pennsylvania 16802, USA}
\affiliation[3]{Center for Multimessenger Astrophysics, The Pennsylvania State University, University Park, Pennsylvania 16802, USA}
\affiliation[4]{School of Natural Sciences, Institute for Advanced Study, Princeton, New Jersey 08540, USA}
\affiliation[5]{Center for Gravitational Physics, Yukawa Institute for Theoretical Physics, Kyoto, Kyoto 606-8502, Japan}

\emailAdd{jac866@psu.edu}
\emailAdd{murase@psu.edu}

\abstract{
New neutrino interactions beyond the Standard Model (BSM) have been of much interest in not only particle physics but also cosmology and astroparticle physics. We numerically investigate the time delay distribution of astrophysical neutrinos that interact with the cosmic neutrino background. Using the Monte Carlo method, we develop a framework that enables us to simulate the time-dependent energy spectra of high-energy neutrinos that experience even multiple scatterings en route and to handle the sharp increase in the cross section at the resonance energy. As an example, we focus on the case of secret neutrino interactions with a scalar mediator.
While we find the excellent agreement between analytical and simulation results for small optical depths, our simulations enable us to study optically thick cases that are not described by the simplest analytic estimates. 
Our simulations are used to understand effects of cosmological redshifts, neutrino spectra and flavors. 
The developments will be useful for probing BSM neutrino interactions with not only current neutrino detectors such as IceCube and Super-Kamiokande but also future neutrino detectors such as IceCube-Gen2 and Hyper-Kamiokande. 
}

\maketitle
\flushbottom

\section{Introduction}
In the Standard Model (SM), astrophysical neutrinos hardly interact as they propagate in intergalactic space from the source to Earth. In SM extensions, $\nu$SI may arise in particle physics models to generate neutrino masses through symmetry breaking \cite{GELMINI1981411,Georgi:1981pg,Chikashige:1980,AULAKH1982136,Davoudiasl:2005ks,Blum:2014ewa,Berryman:2022hds},
which predict new neutrino-neutrino scatterings with the cosmic neutrino background (C$\nu$B), modifying the neutrino energy spectrum in transit. Such secret neutrino interactions, or neutrino self-interactions ($\nu$SIs), are a topic of recent interest in cosmology, due to their effects on the free-streaming behavior of neutrinos and big bang nucleosynthesis
\cite{Hannestad:2004qu,Hannestad:2005ex,Bell:2005dr,Friedland:2007vv,Lancaster:2017ksf,Huang:2017egl,Oldengott:2017fhy,Chu:2018gxk,He:2020zns,Blinov:2019gcj,Kreisch:2019yzn,Brinckmann:2020bcn,Mazumdar:2020ibx,Venzor:2020ova,Huang:2021dba,Forastieri:2019cuf}, which could alleviate the Hubble tension \cite{Blum:2014ewa,Planck:2018vyg,Fields:2019pfx,Araki:2021xdk,He:2020zns,Berbig:2020wve,Berryman:2022hds}. 
These new interactions can modify astrophysical neutrino spectra during propagation
\cite{Ioka:2014kca,Ng:2014pca,Blum:2014ewa,Ibe:2014pja,Kamada:2015era,Shoemaker:2015qul,Murase:2019xqi,Bustamante:2020mep,Creque-Sarbinowski:2020qhz,Esteban:2021tub,Carpio:2021jhu}, as well as supernova neutrino spectra
\cite{Kolb:1987qy,Farzan:2002wx,Blennow:2008er,Farzan:2014gza,Das:2017iuj,Shalgar:2019rqe}.
On the other hand, $\nu$SIs have been constrained by terrestrial experiments, such as $Z$ and $\tau$ decays \cite{Brdar:2020nbj,Berryman:2022hds}, neutrinoless double beta decays \cite{Deppisch:2020sqh,Berryman:2022hds}, and meson decays \cite{Blum:2014ewa,Lyu:2020lps}.

Besides modifications to the neutrino spectrum, scatterings with the C$\nu$B delay the neutrino's arrival to Earth caused by the increased path length. While the scattering angle of high-energy neutrinos is small due to the kinematics of high-energy particle scattering, the large distance between the source and observer can cause a measurable difference in arrival times compared to the photon counterpart~\cite{Murase:2019xqi}.  
Time delays in the neutrino arrival can be used to constrain BSM models and may explore the parameter space that has not been covered by terrestrial experiments. 
We have entered the multimessenger astrophysics era, with coincidences such as the high-energy neutrino event IceCube-170922A and the blazar TXS 0506+056 \cite{IceCube:2018dnn,Amon2018}, IceCube-191001A with the tidal disruption event AT2019dsg \cite{Stein:2020xhk}, and IceCube-200530A and AT2019fdr \cite{Reusch:2021ztx}.
By searching for delayed neutrino emission with not only current IceCube and Super-Kamiokande but also future neutrino detectors such as IceCube-Gen2 and Hyper-Kamiokande, we will be able to constrain some models of nonstandard neutrino interactions by means of statistical analyses. 

Here, we study the neutrino time delay distribution in the context of an interaction term of the form $g\bar{\nu}\nu\phi$, between two neutrinos and a scalar boson $\phi$. 
In particular, we perform dedicated Monte Carlo (MC) simulations of neutrino propagation in three dimensions, taking account of the sudden increase in the optical depth when the neutrino energy approaches the resonance region. The ``time-dependent'' energy spectrum is also calculated, showing the flux suppression near the resonance and the neutrino pileup at energies below the resonance energy. 

In Section 2, we explain the simulation setup and discuss relevant neutrino mean free paths for our trials. 
In Section 3, we analyze four cases: scattering in the small optical depth limit, scattering in the large optical depth limit for both zero and finite inelasticities, and scattering over cosmological distances where the redshift is important. 
In Section 4, we simulate neutrino emission from a source at at redshift $z=1$ with an $\varepsilon_\nu^{-2}$ spectrum and discuss flavor dependence in the resulting time delay distribution and neutrino spectrum.

\section{Method}
In this work, we assume that the $\nu-\nu$ scattering is mediated by a scalar boson $\phi$, of mass $m_\phi$. High-energy neutrinos will scatter off the C$\nu$B via $\nu\nu\rightarrow\nu\nu$~\cite{Ioka:2014kca,Ng:2014pca,Blum:2014ewa}. Assuming that neutrinos are Majorana fermions, we consider the effective Lagrangian for one neutrino generation, $\mathcal{L}\supset -\frac{1}{2}g\overline{\nu_L^c}\nu_L\phi + {\rm c.c.}$, where $g$ is the coupling constant. This model is used for its simplicity, as there is only one neutrino mass and allows us to separate neutrino mixing effects from intrinsic 
features of the BSM scattering. The three-generation case is discussed later. While the high energy neutrinos are ultrarelativistic and left-handed, the C$\nu$B kinetic energy is assumed to be lower than the neutrino mass, so these neutrinos are taken as unpolarized and at rest \cite{Long:2014zva}. 

For a target neutrino mass $m_\nu$ and
incident energy $\varepsilon_\nu$, the scattered neutrino energy $\varepsilon_\nu^\prime$ is given by
\begin{equation}
\varepsilon^\prime_\nu = \frac{\varepsilon_\nu}{1+\dfrac{\varepsilon_\nu}{m_\nu}(1-\cos\theta)},
\label{KinematicConstraint}
\end{equation}
where $\theta$ is the scattering angle. 
Focusing on $s-$channel scattering, we have the angular distribution in the cosmic rest frame 
\begin{equation}
        \frac{1}{\sigma_\nu}\frac{d\sigma_\nu}{d\cos\theta} = \frac{\varepsilon_\nu}{m_\nu}\left(1+\dfrac{\varepsilon_\nu}{m_\nu}(1-\cos\theta)\right)^{-2},
        \label{AngularDistribution}
\end{equation}
and the invariant cross section \cite{Blum:2014ewa}
\begin{equation}
\sigma_\nu(\varepsilon_\nu) = \frac{g^4}{32\pi}\frac{s}{(s-m_\phi^2)^2+m_\phi^2\Gamma_\phi^2},
\label{CrossSection}
\end{equation}
where $s=2m_\nu \varepsilon_\nu$ is the total energy in the center of mass frame and $\Gamma_\phi=g^2m_\phi/16\pi$ is the mediator decay width~\footnote{Note that the coefficient becomes $g^4/16\pi$ if one uses $n_\nu=56~{\rm cm}^{-3}$ instead of $n_\nu=112~{\rm cm}^{-3}$, which is also consistent with the cross section for Dirac neutrinos \cite{Ioka:2014kca,Ibe:2014pja}}. Resonance occurs at $\varepsilon_{\rm res} = m_\phi^2/2m_\nu$, where the cross section becomes $\sigma_\nu=8\pi/m_\phi^2$. In general, the neutrino-neutrino cross section also has $t$-channel contributions and an additional $u$-channel term for Majorana neutrinos. However, our applications lie in the regime $g<0.2$, where these terms are subdominant compared to the $s-$channel term \cite{Blum:2014ewa}. 
The energy distribution for $s-$channel scattering $d\sigma_\nu/d\varepsilon_\nu^\prime$ is flat in the cosmic rest frame at $z$, because the angular distribution in the center-of-momentum frame is isotropic in the scalar mediator case.
Given the angular distribution and our interest in neutrinos above 10 TeV, we expect the scattering angles to be of order $\mathcal{O}(10^{-7})$ and below, as is seen from equation \eqref{AngularDistribution} and $(1-\cos\theta)\varepsilon_\nu/m_\nu\sim 1$. 

If neutrinos interact via a vector mediator, the total cross section would only increase by a constant factor. On the other hand, the angular distribution in the center-of-momentum frame is no longer isotropic: the left-handed neutrino is more likely to scatter in the forward direction. For the same $g$ and $D$, more (less) scatterings would take place in the vector (scalar) mediator case, resulting in typically longer delays.

\begin{figure}
    \centering
    \includegraphics[width =0.7\textwidth]{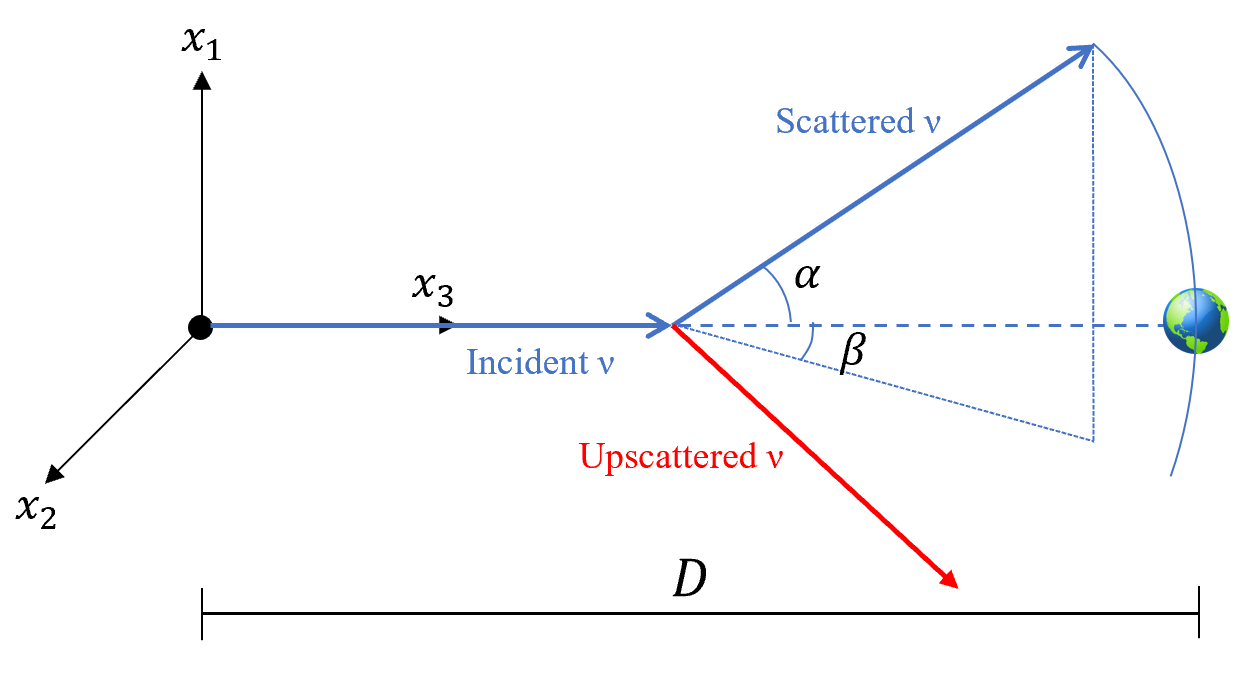}
    \caption{Geometrical setup for our MC simulations. The source is located at the origin, while the observer is at $(0,0,D)$. An outgoing neutrino in the +$x_3$ direction is emitted. Upon scattering, the neutrino is deflected and an additional neutrino is upscattered. Neutrinos stop propagating when they reach the sphere of radius $D$. The angles $\alpha$ and $\beta$ used to compute the time delay are also marked. For illustrative purposes, in this figure the initial neutrino scatters only once.}
    \label{GeometricalSetup}
\end{figure}

In the astrophysical context, time delay due to small-angle scattering was studied e.g., for X-ray scattering \cite{Williamson:1972,Hatchett:1978}, and some techniques are applicable to the current problem of neutrino scattering by using the appropriate differential cross section. The geometrical setup that is used in our simulations will follow that of Ref.~\cite{Williamson:1972}. In Cartesian coordinates, the source is located at the origin, while the observer is at $(0,0,D)$, as shown in figure \ref{GeometricalSetup}. Neutrinos are emitted individually from the source and are tracked until they reach the observer. For a given neutrino path between source and observer, we can make appropriate rotations so that the initial neutrino is always emitted in the $+x_3$ direction, while the final location is some point on the surface of a sphere of radius $D$ with the source as its center. 

Let $\hat{\mathbf{p}}$ be the three-dimensional momentum unit vector of the neutrino and $\alpha$ be the angle that its projection on the $x_1x_3$ plane makes with the $x_3$-axis. Likewise, we define $\beta$ as the angle between the $x_3$-axis and the projection of 
$\hat{\mathbf{p}}$ on the $x_2x_3$ plane. Under the assumption of small-angle scattering, $\alpha,\beta\ll 1$, which applies to our case, 
we neglect terms of third order and higher in $\alpha$ and $\beta$, such that the time delay $t$ of a scattered neutrino compared to an unscattered one is \cite{Williamson:1972}
\begin{equation}
	t = \frac{1}{2}\int_0^D (\alpha^2(x_3)+\beta^2(x_3))dx_3 - \frac{1}{2D}\left[\left(\int_0^D\alpha(x_3) dx_3\right)^2+\left(\int_0^D\beta (x_3) dx_3\right)^2\right].
\label{TimeDelayIntegral}
\end{equation}
The $x_3$ dependence in $\alpha$ and $\beta$ represents the changes in these angles whenever a scattering takes place, thus being applicable for an arbitrary number of scatterings.
Equation~\eqref{TimeDelayIntegral} is evaluated in the MC simulation by splitting 
into a discrete sum, where the steps $dx_3$ correspond to the distance traveled between scatterings. When a scattering takes place, a scattering angle $\theta$ is chosen based on equation
~\eqref{AngularDistribution}, which changes the neutrino's momentum $\mathbf{\hat{p}}$ and hence the values $\alpha$ and $\beta$.
Neutrino propagation stops upon reaching the sphere of radius $D$. 

To determine distances, we choose the cosmological density parameter $\Omega_\Lambda=0.7$, the matter density parameter $\Omega_M=0.3$ and the Hubble constant $H_0=67$ km s$^{-1}$ Mpc$^{-1}$. With these values, a source at redshift $z$ corresponds to a light-travel distance
\begin{equation}
D = \int_0^z \frac{dz^\prime}{H_0(1+z')\sqrt{\Omega_M(1+z^\prime)^3+\Omega_\Lambda}}.
\label{LightTravelDistance}
\end{equation}
We also use this integral to establish a one-to-one correspondence between redshift and neutrino location. 

Let $\varepsilon_\nu$ be the neutrino energy at some redshift. A neutrino initially at position $\mathbf{r}$ may experience a scattering at $\mathbf{r}' = \mathbf{r}+ \hat{\mathbf{p}}dD$ for some traveled distance $dD$ and $\mathbf{r}=(x_1,x_2,x_3)$. To identify $dD$, we also define the optical depth 
\begin{equation}
\tau_\nu = \int_{x_3-dD}^{x_3} n_\nu(x_3^{\prime \prime})\sigma_\nu(\varepsilon_\nu(x_3^{\prime \prime}))dx_3^{\prime \prime},
\label{OpticalDepth}
\end{equation}
where $n_\nu(x_3) = 112$ cm$^{-3}$ $(1+z(x_3))^3$ is the $\nu+\bar{\nu}$ number density of the C$\nu$B and $\varepsilon_\nu$ becomes position dependent as a result of expansion losses. Notice that equation \eqref{OpticalDepth} is a line-of-sight integral and can be used instead of a three-dimensional approach because motion in the other axes is negligible in the small scattering approximation and has little effect in redshift losses.

The probability of an interaction occurring after propagating a distance corresponding to an optical depth $\tau_\nu$ is $1-\exp(-\tau_\nu)$. 
We can thus calculate $dD$ in the MC simulation by drawing $\tau_\nu$ from an exponential distribution and solving equation \eqref{OpticalDepth} for $dD$. 

The main issue when solving for $dD$ is that the cross section can increase by several orders of magnitude as the neutrino energy approaches $\varepsilon_{\rm res}$. The optical depth of a neutrino with energy $\varepsilon_\nu>\varepsilon_{\rm res}$ will then spike as expansion losses cause the neutrino to reach resonance energy. For small $g$, the resonance region is so narrow that a poor choice in $dx_3^{\prime \prime}$ when carrying out the numerical integration of equation \eqref{OpticalDepth} will cause us to miss the resonance entirely.

To tackle this problem, we tabulate the cross section over as a function of the node energy in the range $[\epsilon_{0},\epsilon_{N}]$, for some number of bins $N$, which contains 
$\varepsilon_{\rm res}$ 
and choose a node $k$ such that $\epsilon_k=\varepsilon_{\rm res}$. We consider $k=N/2$ or the integer closest to $N/2$.
We find the nearest resonance at $\epsilon_k$, and the cross section decreases as we move away from $\epsilon_k$.
With $\epsilon_0$ and $\epsilon_N$ fixed, given that $\epsilon_k$ is determined, we then find the value of $\epsilon_i$ that satisfies
\begin{equation}
	\sigma_\nu(\epsilon_i) = \left(\frac{\sigma_\nu(\epsilon_0)}{\sigma_\nu(\epsilon_k)}\right)^{i/k}\sigma_\nu(\epsilon_0), \;\; i\leq k,
\end{equation}
and
\begin{equation}
	\sigma_\nu(\epsilon_i) = \left(\frac{\sigma_\nu(\epsilon_k)}{\sigma_\nu(\epsilon_i)}\right)^{(N-i)/(N-k)}\sigma_\nu(\epsilon_N), \;\; k<i\leq N.
\end{equation}
With this method, we get a larger bin density near resonance as we increase $N$. In this work, we choose $\epsilon_0 = 1$ GeV and $\epsilon_N=10^9$ GeV

We now proceed to outline the method to determine $dD$. We draw a random number by setting $\tau_\nu=-\ln u$ for a random number $u$ uniformly distributed in (0,1]. 
Let $\mathbf{r}$ be the position of our neutrino with energy $\varepsilon_\nu$. As the particle propagates in steps $dx_3^{\prime\prime}$, it accumulates contributions to the optical depth integral $\overline{\tau}_\nu$, following equation \eqref{OpticalDepth} and computed via the trapezium rule. Thus, $dD$ becomes the sum of steps $dx_3^{\prime\prime}$ required to 
to make $\overline{\tau}_\nu = \tau_\nu$. As for the choice of the spacing $dx_3^{\prime\prime}$ used for each contribution to $\overline{\tau}_\nu$, we use the energy nodes $\epsilon_i$ to 
account for redshift energy losses. For the first $dx_3^{\prime\prime}$ we first identify the node $\epsilon_i$ closest to $\varepsilon_\nu$ with $\varepsilon_\nu\geq \epsilon_i$. $dx_3^{\prime \prime}$ is the distance required so that $\varepsilon_\nu$ decreases to $\epsilon_i$ as a result of redshift energy losses alone. 
The next step $dx_3^{\prime \prime}$ is then chosen so that redshift reduces neutrino energy from $\epsilon_i$ to $\epsilon_{i-1}$. Each step calculated via this method keeps increasing the value of $\overline{\tau}_\nu$ and this process is repeated until either reach the sampled $\tau_\nu$ or exceed it. 
If $\overline{\tau}_\nu\geq\tau_\nu$, we interpolate to $\tau_\nu$ and find its associated step $dx_3^{\prime \prime}$. 
It is possible that $\overline{\tau}_\nu\leq\tau_\nu$ throughout the remaining propagation length, in which case the particle is tracked up to the sphere of radius $D$ without further scatterings.
\begin{figure}
    \centering
    \includegraphics[width=0.7\textwidth]{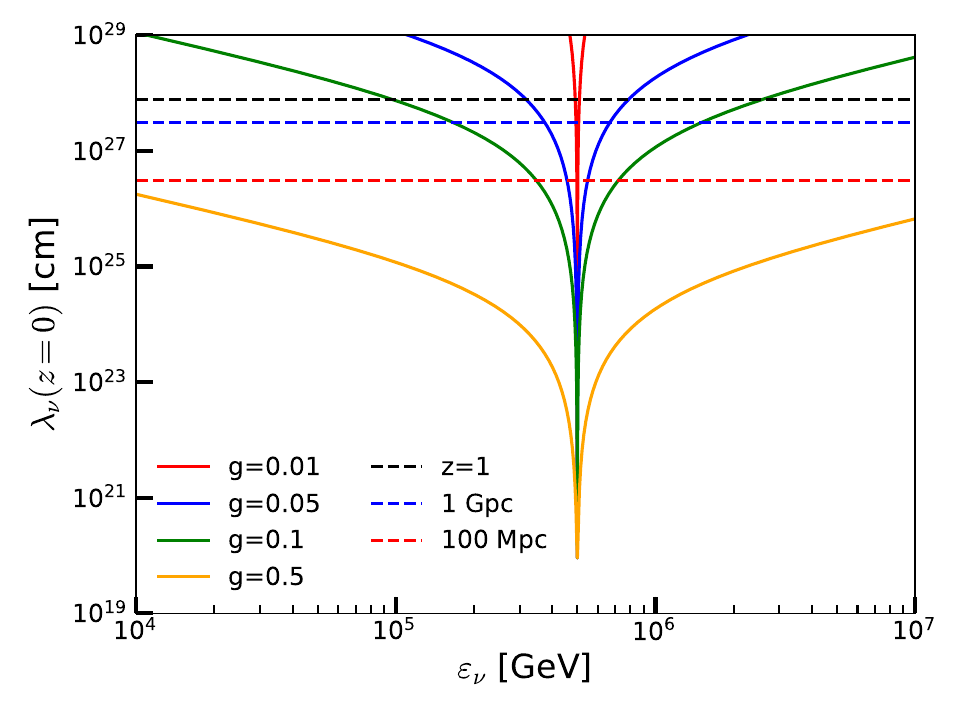}
    \caption{Neutrino mean free path $\lambda_\nu$, at redshift $z=0$, as a function of neutrino energy. We set the parameters $m_\nu=0.1$~eV, $m_\phi=10$~MeV, and choose a variety of coupling parameters $g$. As a reference, we use the light travel distances corresponding to 100 Mpc, 1 Gpc and $z=1$.}
    \label{InteractionLength}
\end{figure}

With $dD$ determined, the particle is moved from $\mathbf{r}$ to $\mathbf{r}'$, the contributions of $\alpha$ and $\beta$ to the integrals $\int\alpha^2dx_3$, 
$\int\beta^2dx_3$, $\int\alpha dx_3$ and $\int\beta dx_3$ in equation~\eqref{TimeDelayIntegral} are computed, and the neutrino energy is redshifted to $\varepsilon_\nu'$ to account for the new position. To perform a scattering, we pick the scattered neutrino energy from a uniform distribution in the 
interval [0,$\varepsilon_\nu'$], since the scattered energy distribution is flat in the cosmic rest frame. From the scattered energy, we can determine the momentum four-vector for both the scattered and upscattered neutrinos, and the upscattered neutrino is injected at $\mathbf{r}'$. 

To find time delay distributions in our examples, we inject neutrinos until the observer collects $10^7$ neutrinos. The energy threshold, below which we do not collect particles, is specified in each example as $E_{\rm th}$.

Henceforth, we choose the values of $m_\phi = 10$ MeV and $m_\nu=0.1$ eV, which sets the neutrino resonance energy to $\varepsilon_{\rm res}=500$~TeV in the cosmic rest frame.
In figure~\ref{InteractionLength}, we show the neutrino mean free path, $\lambda_\nu = 1/n_\nu \sigma_\nu(\varepsilon_\nu)$, using the C$\nu$B density at $z=0$, as a function of the neutrino energy $\varepsilon_\nu$. 
We include the light-travel distances corresponding to 100 Mpc, 1 Gpc and $z=1$, which will be used in our examples.
To describe the regimes of interest, we also introduce the inelasticity parameter $y$, where $y=0$ means that the incident neutrino loses no energy after the scattering. We will also make a distinction between the energy $\varepsilon_\nu$ at redshift $z$, which changes due to cosmological redshift, and the observed neutrino energy $E_\nu=\varepsilon_\nu(z=0)$ at $z=0$. In the first cases, where propagation distances are less than 1 Gpc, adiabatic energy losses do not play a significant role and we have $\varepsilon_\nu\approx E_\nu$. The distinction will be necessary in our examples with sources at $z=1$.

\section{Results}
\subsection{Scattering in the optically thin limit}
As the first example, we consider the propagation of neutrinos with an optical depth of $\tau_\nu\ll 1$, corresponding to the optically thin limit. In this regime, neutrinos are unlikely to
scatter more than once and only a fraction $\tau_\nu$ of all neutrino events will experience a scattering. 

Analytically, the time delay $t$ follows, to a good approximation, the distribution,
\begin{equation}
P(t,\varphi;D) = \frac{1}{2t/D+\varphi^2}\frac{1}{\sigma_
\nu}
\left.\frac{d\sigma_
\nu}{d\theta}\right|_{\theta = \varphi + 2 t/(D\varphi)}
\label{AnalyticalFormula_LowTau}
\end{equation}
where $\varphi$ is the arrival angle on Earth, with respect to the direction of the source. See Ref.~\cite{echoana} for the derivation. Integrating over $\varphi$ will yield the delay distribution $P(t)$.  
The characteristic time delay in the optically thin regime is \cite{Murase:2019xqi}
\begin{equation}
\Delta t\approx \frac{1}{2}\frac{\langle\theta^2\rangle}{4}D \simeq 77\;{\rm s} \;\mathit{C}^2\left(\frac{D}{3~{\rm Gpc}}\right)\left(\frac{m_\nu}{0.1~{\rm eV}} \right)\left(\frac{100~{\rm TeV}}{E_\nu}\right),
\label{LowTauEstimate}
\end{equation}
where $\langle\theta^2\rangle$ is the mean squared angular deflection from a
single scattering. The constant $\mathit{C}\sim$ 1 comes from the angular distribution of the interaction and thus depends on the mediator used.
In the case of $s-$channel scattering, we have $\langle\theta^2\rangle=2\mathit{C}^2m_\nu/E_\nu$, with $\mathit{C}=0.62$ for leading scattered neutrinos \cite{Murase:2019xqi,echoana}.
\begin{figure}
\centerline{
\includegraphics[width=0.7\textwidth]{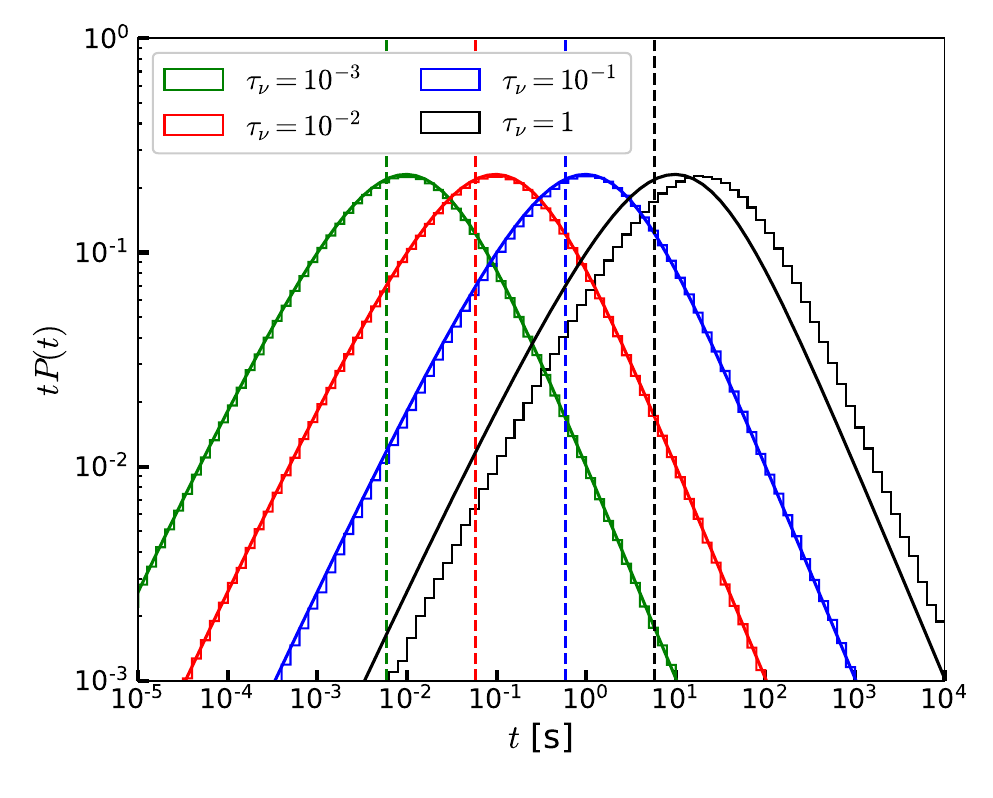}
}
\caption{Time delay probability distribution, for different optical depths with $D=\tau_\nu$~Gpc. The histograms are the results from the MC simulations. The solid curves are obtained by integrating equation
\eqref{AnalyticalFormula_LowTau} over $\varphi$, while the dashed lines are the characteristic time delays given by equation \eqref{LowTauEstimate}.}
\label{TimeDelayDistribution_LowTau}
\end{figure}

To demonstrate our simulation results, we inject neutrinos with $\varepsilon_\nu=170$~TeV and assume $g=0.1$, which leads to $\lambda_\nu = 1$~Gpc. We choose $E_{\rm th}=0$ and construct the time delay distribution $P(t)$, which are shown in figure~\ref{TimeDelayDistribution_LowTau} as histograms for different source distances $D= \tau_\nu\lambda_\nu=\tau_\nu\;{\rm Gpc}$. 
As expected, as $D$ increases, the probability density decreases for shorter $t$. Second, for long time delays we get $P(t)\propto t^{-2}$. 
This is also verified by integrating equation~\ref{AnalyticalFormula_LowTau} over $\varphi$, which is shown as solid curves, and we see the excellent agreement between analytical and numerical results in this optically thin limit. 
The characteristic time delays in equation~\eqref{LowTauEstimate} are also presented as dashed lines. With this example, we also see that our simulation results are consistent with the analytical estimate with leading particles.

We note that for $\tau_\nu\gtrsim0.1$ one can see a visible difference between the numerical and analytical results. At this point, the Poisson probability of two scatterings taking place is $\tau_\nu^2e^{-\tau_\nu}/2$, such that roughly 5\% of the scattered events will scatter twice, causing them to experience longer delays. At $\tau_\nu=1$, the effect of multiple scatterings becomes apparent as we leave the optically thin regime. We note that $\tau_\nu=1$ corresponds to 1~Gpc, where the redshift effect may take place. In this example we ignore redshift energy losses, which will be addressed later.

\subsection{Scattering in the optically thick limit with zero inelasticity}
Let us consider the case where neutrinos do not lose energy, in such a way that the angular distribution in equation \eqref{AngularDistribution} holds but there are no upscattered neutrinos. 
Assuming that multiple scatterings take place, the characteristic neutrino time delay in the large $\tau_\nu$ limit can be estimated as \cite{Murase:2019xqi}
\begin{equation}
    \Delta t \simeq 500\; {\rm s}\; \mathit{C}^2\left(\frac{\tau_\nu}{10}\right)
    \left(\frac{D}{3~{\rm Gpc}}\right)\left(\frac{m_\nu}{0.1~{\rm  eV}}\right)\left(\frac{0.1~{\rm PeV}}{E_\nu}\right)
    \label{HighTauEstimate},
\end{equation}
implying $\Delta t\propto \tau_\nu^2$ for a given $\lambda_\nu$.

The time delay distribution can be expressed as \cite{Williamson:1972,Hatchett:1978}
\begin{equation}
P(t; D) = \frac{4\pi^2}{3\langle\varphi^2\rangle D}\sum_{n=1}^\infty (-1)^{n+1}n^2\exp\left(-\frac{2n^2\pi^2t} {3\langle\varphi^2\rangle D}\right),
\label{AnalyticElasticFormula}
\end{equation}
where $\langle\varphi^2\rangle = \tau_\nu \langle\theta^2\rangle/3$.
When $n_\nu\sigma_\nu t \gg \tau_\nu^2\langle\theta^2\rangle$, which corresponds to long time delays, only the first term of the series is relevant and the probability distribution decreases exponentially. In Ref.~\cite{Williamson:1972}, this distribution is satisfied for the Brownian motion, where $(1/\sigma_\nu)d\sigma_\nu/d\theta$ follows a Gaussian distribution with mean 0 and variance $\langle\theta^2\rangle$. On the other hand, Ref.~\cite{Hatchett:1978}
derives equation~\eqref{AnalyticElasticFormula} under the
assumption that the width of the angular distribution of the
particles in transit is large when compared to the width of the angular distribution of a single scattering (see Ref.~\cite{Hatchett:1978} for details on the assumptions). Our simulations show good agreement with equation~\eqref{AnalyticElasticFormula} when the angular distribution is assumed to follow a Gaussian distribution for $\tau_\nu = 20$ -- $1000$.

\begin{figure}
\centering
\includegraphics[width=0.7\textwidth]{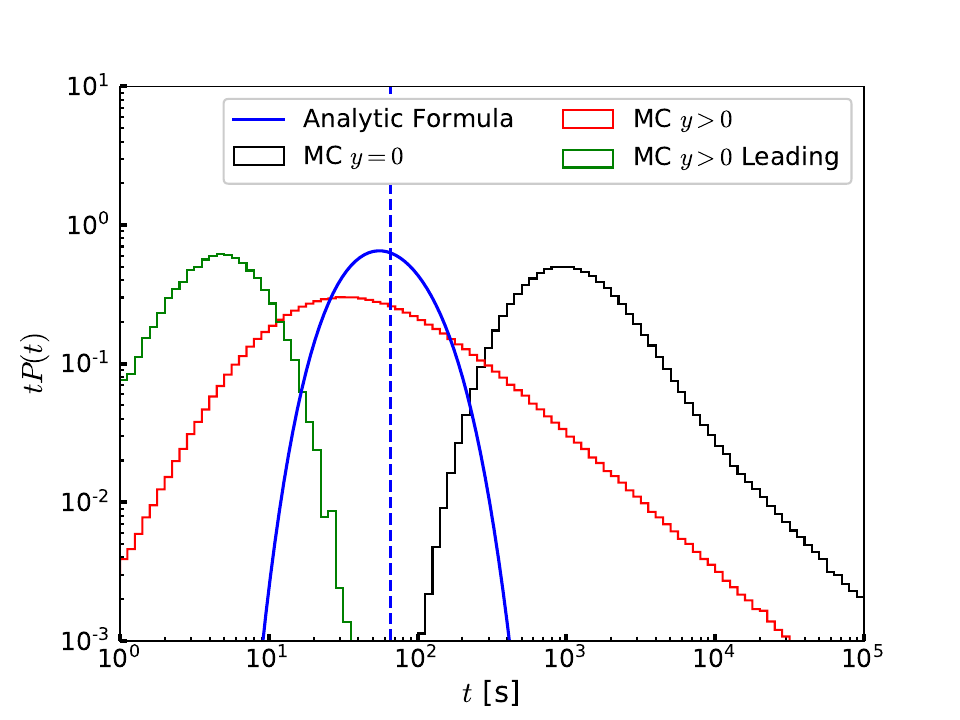}
\caption{Time delay probability density function for the scattering of 300~TeV neutrinos in the $y=0$ regime, with the C$\nu$B at the optical depth $\tau_\nu=310$.
The blue curve is the analytical expression of equation~\eqref{AnalyticElasticFormula}, while the blue dashed line is the typical delay in equation~\eqref{HighTauEstimate}. We also include the results from our MC simulations which do include the finite inelasticity.}
\label{ElasticScatteringComparison}
\end{figure}

In figure~\ref{ElasticScatteringComparison} we show the time delay distribution for 300 TeV neutrinos and $g=0.5$, which gives $\lambda_\nu=10^{24}$~cm. 
We choose $D=100$~Mpc to achieve $\tau_\nu=310$ and the angular distribution used in our simulation follows equation~\eqref{AngularDistribution}. We also compare our result with equation~\eqref{AnalyticElasticFormula} by setting $\langle\theta^2\rangle=0.77m_\nu/E_\nu$, where we assumed $\mathcal{C}=0.62$, and include the typical delay from equation~\eqref{HighTauEstimate}.
We find that the time delay distribution is significantly different from equation~\eqref{AnalyticElasticFormula}: while the analytic expression predicts an exponential decay for large time delays, our simulation suggests that $P(t)\propto t^{-2.1}$. 

We note that in the $y=0$ limit, equation~\eqref{HighTauEstimate} underestimates the characteristic time delay in the sense that the expression relies on $\langle\theta^2\rangle$ to be proportional to the mean number of scatterings $\mathcal{M}=\tau_\nu$, which is true in the case of the Gaussian angular distribution. When we use equation~\eqref{AngularDistribution}, the tail for large $\theta$ is responsible for causing $\langle\theta^2\rangle$ to follow an approximate power law dependence $\tau_\nu^\alpha$ with $\alpha\approx 1.2$, increasing the typical time delay.

As a comparison, we also include the results from the MC simulations with $y>0$, allowing for energy losses and upscatterings of C$\nu$B neutrinos. In this case, the neutrino time delay distribution can be split into the leading and non-leading components. A leading neutrino is ranked based on its energy; at the injection site, the initial neutrino is considered the leading particle. Whenever a leading neutrino scatters, the most energetic of the two outgoing neutrinos is tagged as the leading particle, while 
the other becomes a non-leading particle. In this sense, only a leading neutrino can scatter into another leading neutrino, while non-leading ones remain as such for the duration of the cascade. From this
definition, at any point in the cascade development there can only be one leading neutrino.
The time delays in $y> 0$ are significantly smaller because neutrinos quickly enter the optically thin regime after a few scatterings, so they do not experience $\mathcal{O}(300)$ scatterings as in the $y=0$ case. The leading component has the shortest time delays because the typical scattering angle decreases with $\varepsilon_\nu$. In this case, the analytical expression falls in between the $y=0$ and $y> 0$ regimes.

\subsection{Scattering in the optically thick limit with finite inelasticity}
In realistic scenarios, an incident neutrino loses energy at each scattering, and the energy is transferred to the upscattered neutrino from the C$\nu$B. Multiple scatterings then lead to so-called neutrino cascades \cite{Ioka:2014kca,Ng:2014pca,YOSHIDA1994187}. 
In this example we set $g=0.1$ and $D=500$ Mpc, and look at the scattering of neutrinos with initial energy $\varepsilon_\nu=500$ TeV.

We note that as the incident neutrino loses energy and leaves the resonance region, the cross section will continue to decrease. It is therefore possible that a particle may start off in the optically thick regime, yet ending up in the optically thin regime after a few scatterings, when the mean free path exceeds the propagation length. We can select neutrinos that remain in the optically thick regime by choosing an energy window that is sufficiently close to the resonance, thus avoiding the possibility of a neutrino entering the optically thin regime. 

In the limit that neutrinos cascade down to energies such that the optical depth is below unity, we have the shortest time delay can be estimated with the conservative estimate \cite{Murase:2019xqi},
\begin{equation}
    \Delta t \sim \frac{1}{12}\mathcal{M}\langle\theta^2\rangle\lambda_\nu,
    \label{CascadeEstimate}
\end{equation}
where $\mathcal{M}$ is the mean number of scatterings and can be determined from the MC simulation. 

\begin{figure*}[t]  
    \centerline{
    \includegraphics[width=0.5\textwidth]{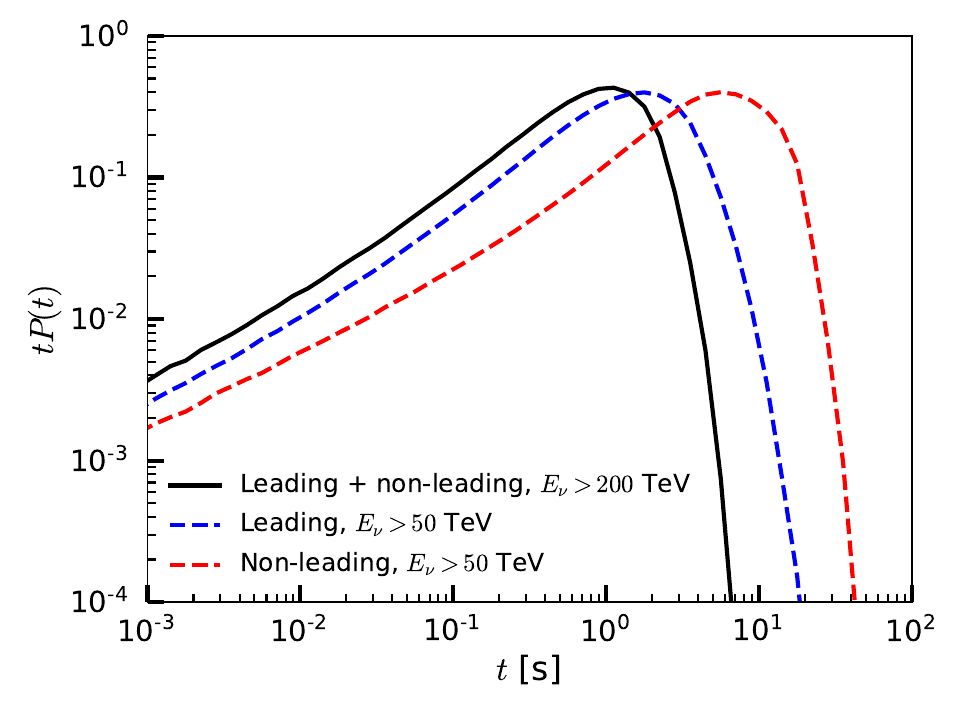}
    \includegraphics[width=0.5\textwidth]{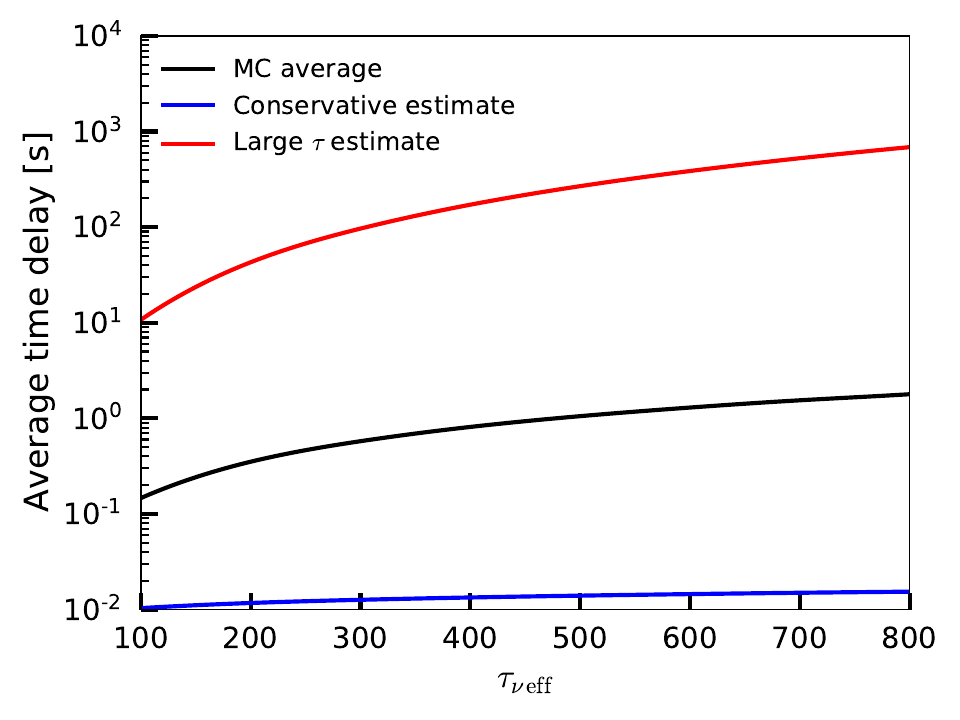}
    }
    \caption{Left panel: Time delay probability distribution, scaled by $t$, for a source distance $D = 500$~Mpc and initial neutrino energy $\varepsilon_\nu=500$~TeV. The distribution of all particles above 200~TeV (mostly leading particles) is shown by the black curve, as well as the leading and non-leading components for neutrinos with energy above 50~TeV, by the blue and red curves respectively. Right panel: MC average time delay of $E_\nu>200$ TeV neutrinos, as a function of the effective optical depth in the 200~TeV -- 500~TeV energy range. This time delay is compared to the large optical depth estimate and the conservative estimates, given by equations~\eqref{HighTauEstimate} and \eqref{CascadeEstimate}, respectively.  }
    \label{InelasticDistribution}
\end{figure*}

To account for energy losses, we define the effective optical depth $\tau_{\nu\,{\rm eff}}$ as the optical depth using the average cross section over the energy window, which we choose as 200 TeV -- 500 TeV. The quantity $\tau_{\nu\,{\rm eff}}$ is defined for illustrative purposes to explain the physics by using a single optical depth and is not used in the simulations themselves.
We show our results in figure~\ref{InelasticDistribution}, where the source distance $D=500$~Mpc corresponds to $\tau_{\nu \,{\rm eff}}=600$. The resulting distribution is shown by the black curve and the neutrinos that generate this distribution are mostly leading particles. By definition there are no non-leading particles with energies above $\varepsilon_{\rm res}/2=250$ TeV, because when the first scattering occurs, only the leading neutrino will have energy above 250 TeV. Any non-leading neutrino from the cascade will never have more than half the energy of the initial neutrino. The drop in the distribution for long time delays is caused by the threshold, as particles with less energies are typically the ones with the larger scattering angles and time delays, by virtue of equation~\eqref{AngularDistribution}.

We also include the leading and non-leading components at energies above 50~TeV. For this threshold we cannot guarantee the optically thick regime, but including these highlight the shift to longer time delays as the energy threshold decreases. As expected, the leading component is associated with shorter time delays when compared to the non-leading.

On the right panel of figure~\ref{InelasticDistribution}, we compare the average time delay with the estimates provided by equations~\eqref{HighTauEstimate} and \eqref{CascadeEstimate}. The time delays are given as a function of the effective optical depth in all three cases, using the energy range 200~TeV -- 500~TeV. Since almost all the neutrinos in this energy range are leading particles, the MC average will not change if we only consider leading neutrinos. For the conservative estimate, we find that $\mathcal{M}$ increases slowly, from 1.8 at $\tau_{\nu\, {\rm eff}}=100$ to 2.7 at $\tau_{\nu\, {\rm eff}}=800$. For $y=0$, we would have $\mathcal{M}\propto\tau_\nu$, but in the presence of energy losses, most of the particles that experience multiple scatterings lie below the threshold and are not counted in the calculation of $\mathcal{M}$.

\subsection{Scattering over cosmological distances}
When the source is located at non-negligible redshifts, we must account for neutrino energy losses due to the expansion of the Universe. Here, we use a coupling constant of $g=0.01$, providing a very small energy window for the neutrino to interact (see figure~\ref{CrossSection}). 

As an example, we consider a neutrino source at $z=1$, which corresponds to a light-travel distance of $D=2.5$ Gpc, emitting 800 TeV neutrinos. Assuming redshift losses only, the neutrino energy reaches $\varepsilon_{\rm res}$ at $z=0.25$. In the vicinity of $z=0.25$, a scattering will take place and the neutrino will then lose energy such that it is no longer in the resonance window.

\begin{figure}
    \centering
    \includegraphics[width=0.7\textwidth]{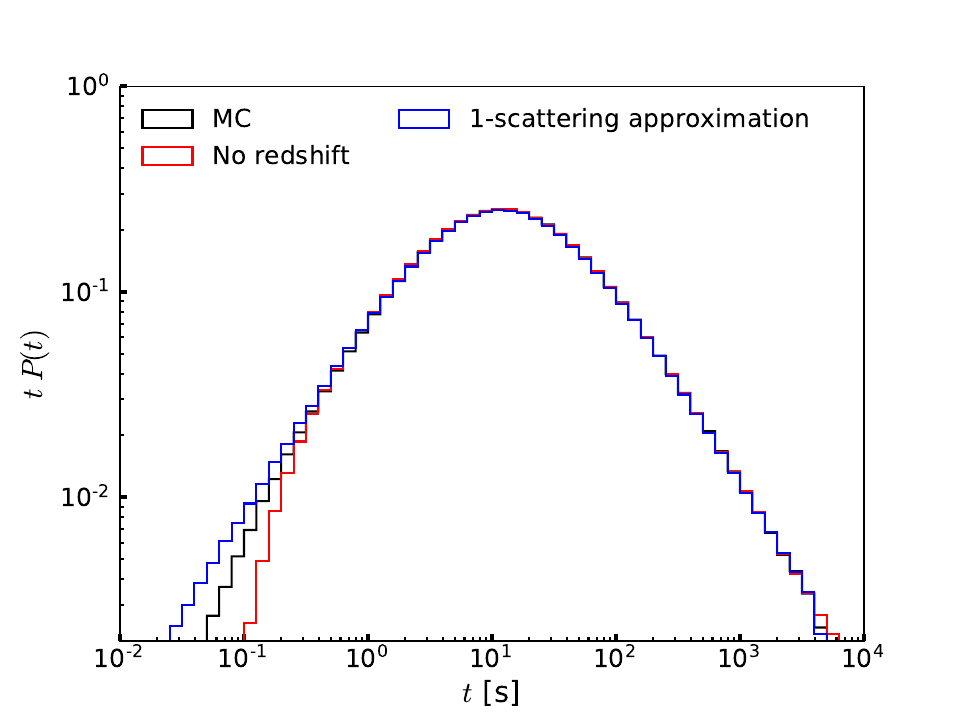}
    \caption{Time delay distribution of 800~TeV neutrinos starting at $z=1$ and scattering off the C$\nu$B. The black histogram is the result from our MC simulation. The red histogram is a separate simulation, where redshift energy losses are ignored and the neutrino's energy is manually changed to $\varepsilon_\nu = \varepsilon_{\rm res}$ at $z=0.25$. The single scattering approximation, which assumes that the cross section is a Dirac delta function spiking at $\varepsilon_{\rm res}$, is shown as the blue histogram.}
    \label{CosmologyExample}
\end{figure}

The resulting time delay distribution is presented in figure~\ref{CosmologyExample}. Together with the MC simulation, we include the case without the redshift effect, where we ignore redshift loss, but manually change the neutrino energy to $\varepsilon_\nu = \varepsilon_{\rm res}$ at $z=0.25$ and allow the particle to scatter the C$\nu
$B. 
The distribution shown by the red curve shows the redshift effect in the transition from the optically thick to the optically thin regime. We also show the case of the single scattering approximation, which treats the cross section as a Dirac delta function that spikes at $\varepsilon_{\rm res}$, and this is represented by the blue curve. 
We see that the single scattering approximation correctly predicts the MC results, except for short time delays of $t<1$ s. In the single scattering approximation, the short time delay portion originates from particles that experience small-angle scatterings and keep their energies very close to $\varepsilon_{\rm res}$ within less than $1\%$. In reality, these particles should scatter again, since they are still within the resonance region. Upon the second scattering, the time delay is expected to increase, which is why the scenario ignoring redshift losses has a deficit in events with $t \lesssim 0.3$~s. If we then compare the red curve to the MC distribution, we see that this deficit is less significant. Once adiabatic energy losses are considered, small changes in $z$ as the neutrino propagates are enough to shift the energy away from the resonance, and increase the mean free path, facilitating the transition to the optically thin regime. 

Our treatment is applicable to cosmological sources as long as particles travel almost along the line of sight. When the scattering angle is not small, due to cosmological expansion, the delay may scale as, e.g., $\propto(1+z_{\rm sc})l_{\rm prop}$ instead of the light-travel distance, where $l_{\rm prop}$ is the proper scattering length and $z_{\rm sc}$ is the redshift where the scattering occurs. However, this effect would increase the delay by $\sim 10\%$ for $z\lesssim 1$. 

\section{Applications}
\subsection{Source spectra}
In the previous section, we have focused on monoenergetic spectra at the source. Here, we analyze effects of neutrino-neutrino scattering assuming an $\varepsilon_\nu^{-2}$ power law spectrum from a source at redshift $z=1$ and set a threshold energy of 1 TeV.

\begin{figure}
    \centering
    \includegraphics[width=0.9\textwidth]{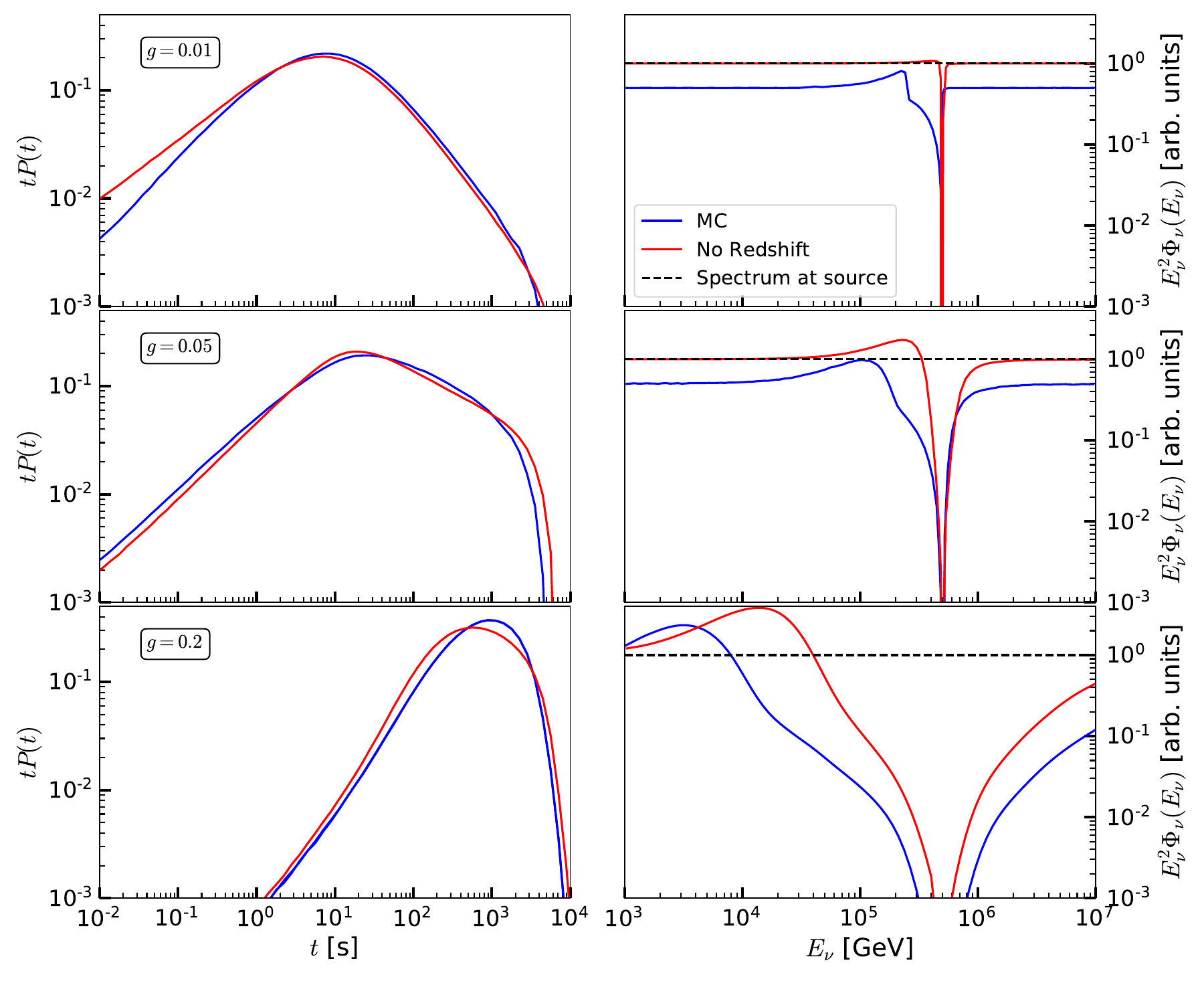}
    \caption{Time delay distributions (left panels) and observed energy spectra (right panels) of an $\varepsilon_\nu^{-2}$ source at $z=1$, for coupling constants $g=0.01,0.05$ and 0.2 (top,middle and bottom row, respectively). The source spectrum is normalized such that $\varepsilon_\nu^2\Phi_\nu=1$. The blue curves represent the results of our MC simulation, while the red curves correspond to a case where the redshift energy loss and C$\nu$B density dependence on $z$ are neglected. The dotted black lines in the energy spectra are the neutrino spectra at the source.}
    \label{E-2Spectrum}
\end{figure}

As examples, we consider values of the coupling, $g=0.01,0.05$ and 0.2, as they represent the $\tau_\nu\ll 1, \tau_\nu\sim 1$ and $\tau_\nu\gg 1$ regimes in the 100~TeV--1~PeV range, as shown in Figure
\ref{E-2Spectrum}. The results from the MC simulations are shown by the blue curves, while separate simulations ignoring redshift effects, meaning no expansion losses and assuming a uniform C$\nu$B 
number density of 112~cm$^{-3}$, are shown by the red curves.

Starting with $g=0.01$, we see that the time delay distribution close to the $t P(t)$ peak is not very sensitive to redshift effects and $P(t)\propto t^{-2}$ past the peak. Below the peak, we see there are more events  with $t< 1$ s when we neglect redshifts. 
In the absence of redshifts there is a sharp decrease at 500 TeV due to the resonance, together with the corresponding pileup in the 400 TeV region. 
This occurs over a very narrow energy region, and the pileup is not very significant because few neutrinos lie in the resonance windows. In the realistic scenario we see the expected decrease in the normalization, with $E_\nu^2\Phi_\nu$ scaling as $1/(1+z)$. The pileup region shifts towards lower energies, and the peak is more prominent. 
Because neutrinos from the higher-energy tail get redshifted into the resonance region and scatter, it follows that the total number of scattered neutrinos in the presence redshift is larger than the case without. There is also a distinct break in the spectrum at the 250 TeV mark, which is understood by differentiating between scattered and unscattered neutrinos. The component of the initial $\varepsilon_\nu^{-2}$ spectrum that was below 500 TeV remains unscattered and is simply redshifted to 250 TeV and below. On the other hand, neutrinos between 500 TeV and 1 PeV will eventually scatter as they get redshifted into the resonance window, while those above 1 PeV are redshifted to a minimum energy of 500 TeV and do not interact. Therefore, the observed energy spectrum of unscattered neutrinos is an $\varepsilon_\nu^{-2}$ spectrum with a gap in the 250 TeV-- 500 TeV region, which is to be filled by the scattered component. For couplings this small, there are not enough upscattered neutrinos to cover this gap, causing the discontinuity.

For $g=0.05$ we have multiple scatterings, typically four to five, which causes the $t P(t)$ peak to appear at $t \approx 20$~s. For long time delays, we also see a sudden drop around 5000~s. This is caused by the 1~TeV energy threshold, which removes lower-energy neutrinos that would have a larger scattering angle and longer time delay. The energy spectrum shows features similar to $g=0.01$, but the pileup region is wider as a result of multiple scatterings. The spectrum between 100 and 500 TeV in the MC case is dominated by the scattered component, so we no longer see the break in the spectrum observed when $g=0.01$.

The case where $g=0.2$ shows a large separation in the time delay distribution peaks between the redshift and no redshift cases. Here, the number of neutrino scatterings is much higher, many of them experiencing over 15 scatterings. In this case, redshift losses decrease the neutrino energy before the next scattering takes place, at which point larger scattering angles are preferred, according to equation \eqref{AngularDistribution}. The threshold effect on the delay distribution occurs close to $10^4$~s, but does not have an effect on the location of the distribution peaks, which is also true for $g=0.05$ and $g=0.01$. Setting the energy threshold to 10~TeV, however, would shift the peak locations to lower $t$ for $g=0.2$ only. 
Looking at the energy spectrum, we see that the pileup region is much wider. 
We also note that $E_\nu^2\Phi_\nu$ is no longer flat in the 1 TeV region and the MC result overcomes the no redshift case at low energies. 
Now that the number of scatterings is so large, particle multiplicity allows the MC peak to compensate for the redshift factor $1/(1+z)$. On the other hand, there is a drop in the case without the redshift effect, because TeV neutrinos experience scatterings at such large couplings, and the higher-energy neutrinos that cascade downward are unable to compensate. After repeating these simulations with the inclusion of the $t-$channel contributions to the cross section, we find negligible differences for $g=0.01$ and $g=0.05$ and a slight shift to longer time delays for $g=0.2$. The differences only appear at large couplings, where the resonance width is large and the $t-$channel term is comparable to the $s-$channel away from $\varepsilon_\text{res}$.

\subsection{Flavors}
If we consider three neutrino flavors, the cross section has to be modified for different mass eigenstates $m_i$. The oscillation parameters are fixed to the best-fit oscillation results from NuFIT 2021 \cite{Esteban:2020cvm,NuFit2021}. While there are three mass eigenstates, neutrino oscillation data tell us that two of these are close together. We should then expect two well-separated resonance dips. To comply with the cosmological bound of $\sum m_\nu<0.12$ eV \cite{Planck:2018vyg}, we choose the masses $m_1=0.022$~eV, $m_2=0.024$~eV and $m_3=0.055$~eV. In addition, to obtain the dips in the energy spectrum between 100~TeV and 1~PeV, we choose $m_\phi = 5$~MeV.

The C$\nu$B density for each mass eigenstate is $112(1+z)^3$ cm$^{-3}$, as before. Regarding the $\varepsilon_\nu^{-2}$ source at $z=1$, we will assume that the
flavor ratio at the source is (1:2:0), which quickly decoheres into mass eigenstates as the neutrino oscillation and coherence lengths are shorter than the interaction length. The propagation and interactions can thus be carried out in the mass eigenstate basis and then converted to the flavor eigenstate basis when it reaches the source. 

The neutrino coupling now becomes a $3\times 3$ coupling matrix, and we assume the coupling only for $\nu_\tau$: $g_{\alpha\beta} ={\rm diag}(0,0,g_{\tau\tau})$. 
Such secret neutrino interactions involving only $\nu_\tau$ are of interest as they are the least constrained by laboratory experiments \cite{Blinov:2019gcj}. 
Under this assumption, the invariant cross section for the process $\nu_i\nu_j\longrightarrow \nu_k\nu_l$ is \cite{Blum:2014ewa}
\begin{equation}
\sigma_{\nu}^{ijkl}= \frac{|U_{\tau i}|^2|U_{\tau j}|^2|U_{\tau k}|^2 |U_{\tau l}|^2g_{\tau\tau}^4}{32\pi}\frac{s_j}{(s_j-m_\phi^2)^2+m_\phi^2\Gamma_\phi^2},
\label{CrossSection3Flavors}
\end{equation}
where $s_j = 2m_jE_\nu$, $\Gamma_\phi=g_{\tau\tau}^2m_\phi/16\pi$ and $U$ is the Pontecorvo-Maki-Nakagawa-Sakata matrix.

\begin{figure*}[t]
    \centerline{
    \includegraphics[width=0.5\textwidth]{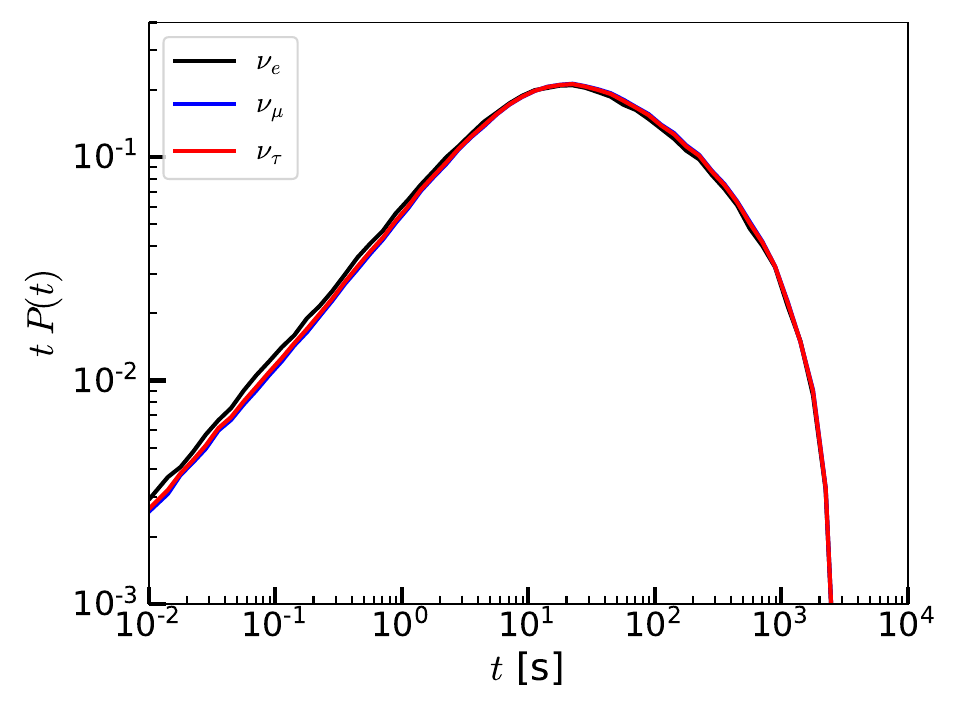}
    \includegraphics[width=0.5\textwidth]{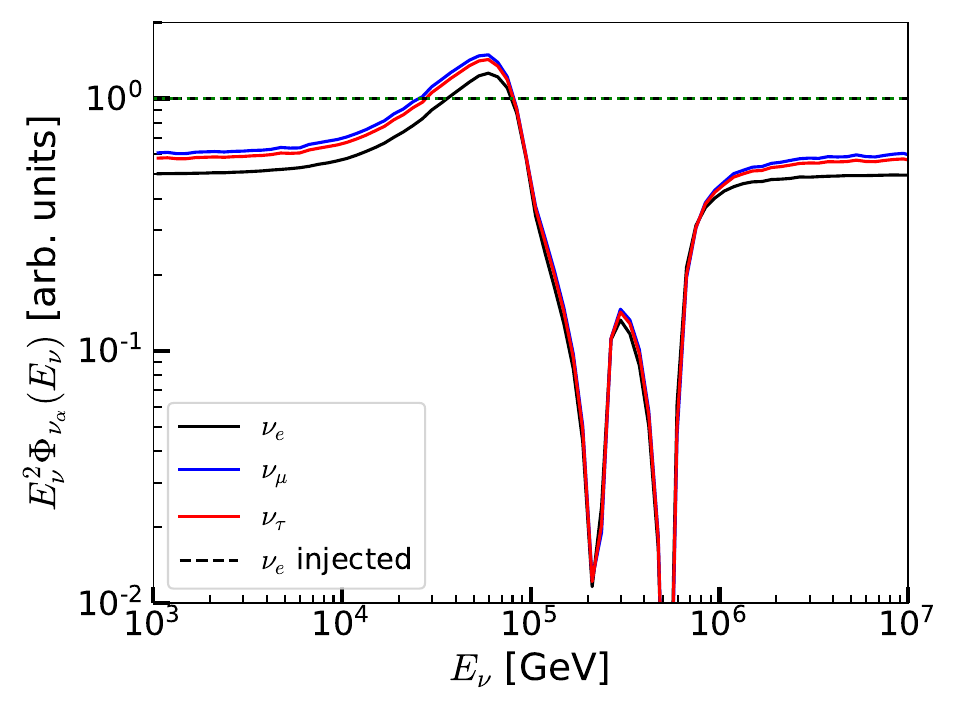}
    }
    \caption{Time delay distributions (left panel) and observed energy spectra (right panel) of an $\varepsilon_\nu^{-2}$ source at $z=1$, for $g_{\tau\tau}=0.05$ and $m_\phi = 5$~MeV. The source spectrum is normalized such that $\varepsilon_\nu^2\Phi_{\nu_e}=1$. }
    \label{NuTauOnly}
\end{figure*}

The results of the MC simulation are shown in figure~\ref{NuTauOnly} for $g_{\tau\tau}=0.05$. The energy spectrum shows two dips due to the three resonances. Besides that, the spectral shape is about the same for all three neutrino flavors, separated by factors which correspond to the observed flavor ratio at Earth after oscillations are averaged out. 
When a scattering takes place, the outgoing mass eigenstates $\nu_k$ and $\nu_l$ depend on $|U_{\tau k}|^2$ and $|U_{\tau l}|^2$ only. For our choice of oscillation parameters, we have $|U_{\tau 1}|^2<|U_{\tau 2}|^2<|U_{\tau 3}|^2$. As a result, there is a slight tendency for $\nu_3$ to be produced over the other states, which builds up over several scatterings, creating the deficit in $\nu_e$ when we convert the $\nu_3$ flux to a flavor flux.
Our results on the spectra are consistent with those by Ref.~\cite{Blum:2014ewa}. 

For the time delay, the delay distributions are almost identical. One could see that $P(t)$ is slightly larger for $\nu_e$ in the 0.01~s -- 1~s range. This part of the distribution comes from neutrinos that only scatter once, while the long time-delay tail consists of particles that scatter multiple times. 

\section{Summary and Conclusions}
We have presented a numerical study of secret neutrino interactions of TeV--PeV neutrinos and their associated time delays. We developed a MC simulation code that accounts for the sudden changes in the $s$-channel interactions as we approach the resonance energy, allowing us to accurately calculate the scattering locations. The developments can be applied to various astrophysical neutrino sources, by which constraints on $\nu$SI can be placed with neutrino data (see Ref.~\cite{echoSN} as an application to the Galactic supernova).  

As the first example, we have shown that in the optically thin limit the simulation result is in agreement with the analytical expression. Deviations from it become apparent at $\tau_\nu\sim 0.1$, when multiple scatterings become more relevant. In the optically thick limit with $y=0$, there is a significant difference in the time delay distribution between our result from the MC simulation and the analytical expression, because the angular distribution is not a Gaussian. 
The case $y=0$ predicts longer delays than $y>0$ as energy losses allow particles to leave the resonance window, causing less scatterings to take place. In the case $\tau_\nu\gg 1$ and $y>0$, we have found that the time delay distribution is also sensitive to the energy threshold: lowering it leads to the inclusion of the lower-energy particles that experience more scatterings and longer time delays. A clear separation between the distribution peaks for leading and non-leading components is seen at an energy threshold of 50 TeV. The characteristic time delays in the MC simulations are found to lie between the large optical depth and conservative estimates. 

Considering sources at cosmological distances, we have shown that for a source at $z=1$ and the coupling strength $g=0.01$, redshift effects are most important for neutrinos in the short time-delay tail. We have also highlighted the difference between the MC simulation and the single scattering approximation, and the latter predicts more events in the short time-delay tail, compared to the former. 
For a source at redshift $z=1$ with an $\varepsilon_\nu^{-2}$ spectrum, the observed neutrino spectrum presents the expected pileup region below the resonance energy. As the coupling strength increases, the resonance width increases and the location of the pileup moves to lower energies. Breaks in the spectrum at coupling strengths $g=0.01$ and $g=0.05$ are present, at energies slightly above the pileup region, where the scattered and unscattered components of the spectrum intersect. This effect is not present when we ignore redshift effects. The time delay distributions for larger couplings lead to longer delays, as more scatterings occur, with delays of approximately 1~hour for $g=0.05$.

The MC simulation code developed in this work can also be applied for a broader set of BSM interactions. As long as the small-angle scattering approximation is satisfied, then the MC code presented here can be applied to neutrino scattering with dark matter or axions, as discussed in Refs.~\cite{Murase:2019xqi,Koren:2019wwi}. 
Other BSM interactions which produce SM particles such as muons and pions, which decay into neutrinos, can also be accommodated readily. The code is expected to be publicly available in the near future.

\acknowledgments
We thank Ali Kheirandish and Shigeru Yoshida for useful discussions and comments. 
The work of K.M. is supported by the NSF Grant No.~AST-1908689, No.~AST-2108466 and No.~AST-2108467, and KAKENHI No.~20H01901 and No.~20H05852. J.C. is supported by the NSF Grant No.~AST-1908689.

\bibliographystyle{JHEP}
\bibliography{bibtex}

\end{document}